\documentclass[showpacs, pra,twocolumn,preprintnumbers ,amsmath, amssymb, superscriptaddress, aps]{revtex4-2}
\usepackage{color}
\usepackage{amsmath,amssymb}
\usepackage{pifont}% http://ctan.org/pkg/pifont
\usepackage{amssymb}  % More symbols
\usepackage{bbold}
\usepackage{float}
\usepackage{subfloat}
\usepackage[squaren]{SIunits}

\usepackage[caption=false]{subfig}
\usepackage{tikz}
\usepackage{makecell}
\usepackage{subfig}
\usepackage{pifont}   % Ding symbols
\usepackage{graphicx} % Include figure files
\graphicspath{{Figures/}}
\usepackage{dcolumn}  % Align table columns on decimal point
\usepackage{bm}       % bold math
\usepackage{multirow} % Table functions
\usepackage{placeins}% For making floats not move around everywhere
\usepackage[colorlinks]{hyperref}
\usepackage{mathtools}
\usepackage{appendix}

\begin{document}
	
	\title{
Controlling Goos-Hänchen shifts in phosphorene via barrier and well}
	\date{\today}
	
	\author{Jilali Seffadi}
		\affiliation{Laboratory of Theoretical Physics, Faculty of Sciences, Choua\"ib Doukkali University, PO Box 20, 24000 El Jadida, Morocco}
      	\author{Hocine Bahlouli}
      \affiliation{Physics Department 5047, King Fahd University of Petroleum and Minerals, Dhahran 31261, Saudi Arabia}  
	\author{Ahmed Jellal}
	\email{a.jellal@ucd.ac.ma}
	\affiliation{Laboratory of Theoretical Physics, Faculty of Sciences, Choua\"ib Doukkali University, PO Box 20, 24000 El Jadida, Morocco}
	\affiliation{Canadian Quantum  Research Center,
		204-3002 32 Ave Vernon,  BC V1T 2L7,  Canada}


		\begin{abstract}

{ We study the effect of the double potentials (barrier, well) on the Goos-Hänchen (GH) shifts in phosphorene. We determine the solutions of the energy spectrum associated with the five regions that make up our system. By studying the phase shifts, we find that the GH shifts are highly sensitive to the incident energy, the $y$ directional wave vector, the potential heights and widths. To validate our findings, we perform a numerical analysis of the GH shifts as a function of the transmission probability under various conditions. In particular, we observe a consistent pattern in which a positive peak in the GH shift is always followed by a negative valley, a behavior evident at all potential height values. Notably, the energies at which the GH shift changes sign coincide exactly with the points at which transmission drops to zero. In particular, the transmission resonances that occur just before and just after the transmission gap region are strongly correlated with the points at which the GH shift changes sign. This study advances our understanding of how the double potential influences the GH shift behaviors in phosphorene. The ability to fine-tune the GH shifts by changing system parameters suggests potential applications in optical and electronic devices using this two-dimensional material.}
			
		\end{abstract}

	\maketitle

\section{Introduction}
A monolayer of black phosphorus, called phosphorene, can be stripped down to extremely thin layers. The phosphorus (P) atoms are arranged in a staggered hexagonal lattice. It is a layered material, similar to graphite and transition metal dichalcogenides (TMDs), where each atomic layer is bound to adjacent ones by weak van der Waals interactions, allowing the isolation of a few atomic layer thin films by mechanical exfoliation \cite{ref-1,ref-2,ref-3}. The intrinsic anisotropy of phosphorene is observed in its optical \cite{Wein2014}, thermal \cite{Quin2015}, electrical \cite{Li2014} and mechanical \cite{Liu2014} properties , making it an intriguing material with potential applications in devices based on thermoelectronic, spintronic and electronic effects.

{In contrast to graphene}, phosphorene exhibits an intrinsic direct band gap that is layer dependent and varies between $0.3$ eV (for bulk) and $2$ eV (for single layer) \cite{ref-7}. It has also been observed that the layer thickness of phosphorene has a significant effect on the time response ratios for ON/OFF, OFF, and ON currents in phosphorene field effect transistors (FETs) \cite{ref-2}. {Note that as far as TDMs are concerned, the experimentally measured band gaps are generally in the range of 1.87-1.92 eV for MoS$_2$ \cite{Saadati2021}, 1.55-1.58 eV for MoSe$_2$ \cite{Wang2012}, 1.98-2.05 eV for WS$_2$, and 1.6-1.66 eV for WSe$_2$ \cite{Gusakova2017}.}

Recently, significant progress has been made in the study of electronic transport properties in graphene systems \cite{100,Akhavan2010, Bai2010}. One example is the quantum version of the Goos-Hänchen (GH) effect, which arises from the complete reflection of particles at the interfaces between two different media. It was first observed by Hermann Fritz Gustav Goos and Hilda Hänchen \cite{200,300}, the GH shifts were theoretically explained by Artman in the late 1940s \cite{400}. Several studies focusing on different graphene-based nanostructures, such as single barrier \cite{500}, double barrier \cite{600}, and superlattices \cite{700}, have shown that GH shifts can be enhanced by transmission resonances. In addition, these shifts can be controlled by adjusting the electrostatic potential and the induced gap \cite{500}. Similar to semiconductors, electrical and magnetic barriers can modulate GH shifts in graphene, similar to observations in atomic optics \cite{800,900}. Crucially, research has highlighted the central role of GH shifts in determining the group velocity of quasiparticles along the interfaces of graphene p-n junctions \cite{1000,1100}. Furthermore, investigations have shown that the valley-dependent GH effect in strained graphene leads to different group velocities for electrons near the valleys $K$ and $K'$ \cite{1200}.

{In \cite{ref92}, we have studied the transport properties of charge carriers in phosphorene subjected to two potentials: barrier and well. The influence of the potential parameters and the wave vector of the incident particles in the $x$-direction on the energy spectrum is studied. Using appropriate boundary conditions and the transfer matrix method, we have calculated the transmission coefficient and the conductance. These values were thoroughly analyzed in order to discern the fundamental properties of our system, including their dependence on physical parameters and the armchair boundary condition. Our results provided strong evidence for the pronounced anisotropy of phosphorene and verified the occurrence of Klein tunneling at normal incidence. Furthermore, under certain conditions, we observed oscillatory patterns in transmission and conductance with respect to the potential widths.}

{The GH effect in phosphorene subjected to one and two potential barriers is studied in \cite{Majari2023}, where the basic concepts and theoretical framework for discussing G-H shifts in the context of different types of barriers are introduced. Building on this foundation, our work specifically investigates the effect of a barrier and a well, thereby broadening the scope of the study. More specifically, we introduce further complexity by analyzing how the GH shifts are affected by the heights and widths of double potentials. Our research provides a more sophisticated perspective that extends the ideas presented in \cite{Majari2023}.}

Motivated by the previous works \cite{Majari2023} and \cite{ref92}, we study the effect of double potentials (barrier, well) on the GH shifts in phosphorene, building upon our earlier research \cite{ref92}. Depending on the phase shifts, we have determined the GH shifts, which are highly sensitive to the physical parameters of the system under consideration. To provide a comprehensive analysis of our findings, we present a numerical investigation of the GH shifts associated with the transmission probability under various conditions. More specifically, we demonstrate that the GH shifts can exhibit both negative and positive signs, depending on the choice of parameters. Our results also indicate that the system's characteristics can be leveraged to control the GH shifts in terms of both magnitude and sign. This work expands upon the understanding of how double potentials impact the GH shift behaviors in the Dirac material phosphorene. The ability to tune the GH shifts through the selection of system parameters suggests potential applications in optical and electronic devices based on this two-dimensional material.
 
{The GH shifts in phosphorene can be tuned and controlled, opening up a number of exciting possibilities for applications in various industries. Here are some specific applications where this tuning can be particularly beneficial. Phosphorene can be used in high precision optical devices such as waveguides and beam splitters by varying the GH shifts. This could allow precise control over the propagation of light, which could improve the performance of photonic circuits. Because GH varies in sensitivity to physical conditions, phosphorene is a suitable material for the development of ultrasensitive sensors. Changes in chemical composition, biological interactions, or environmental variables could be detected by these sensors. Modulation of the GH shifts could facilitate the manipulation of quantum states in quantum information systems. Because of the special electrical properties of phosphorene, more effective quantum gates or qubits could be created. Phosphorene can enhance light-matter interactions in integrated photonic circuits by tuning the GH shifts, which could lead to improved functionality in switches and modulators, among other devices. Controlling the GH shifts can be useful for SPR sensors, which depend on precise light management at the material interface. This can improve the sensitivity and specificity of biosensors. By increasing the efficiency of light extraction and absorption, fine-tuning the GH shifts can maximize the performance of optoelectronic devices such as solar cells and light-emitting diodes (LEDs). }

The present paper is organized as follows: We establish our mathematical model and describe the Hamiltonian of the system in Sec. \text{\ref{Sec2}}. Next, we obtain the corresponding energy spectrum of our system by solving our stationary wave equation in each region of space. The associated phase shift is then inferred by computing the transmission probability, and this is used to determine the corresponding GH shifts in Sec. \text{\ref{Sec3}}. We numerically computed the GH shifts and transmission by carefully choosing the physical parameters of the system and discussed the fundamental aspects of our system in Sec. \text{\ref{Sec4}}. Finally, we summarize our results.

%%%%%%%%%%%%%%%%%%%%%%%%%%%%%%%%%%%%%%%%%%%%%%%%%%
\section{Theoretical Model}\label{Sec2}
\begin{figure}[ht]
	\centering
	\includegraphics[scale=0.4]{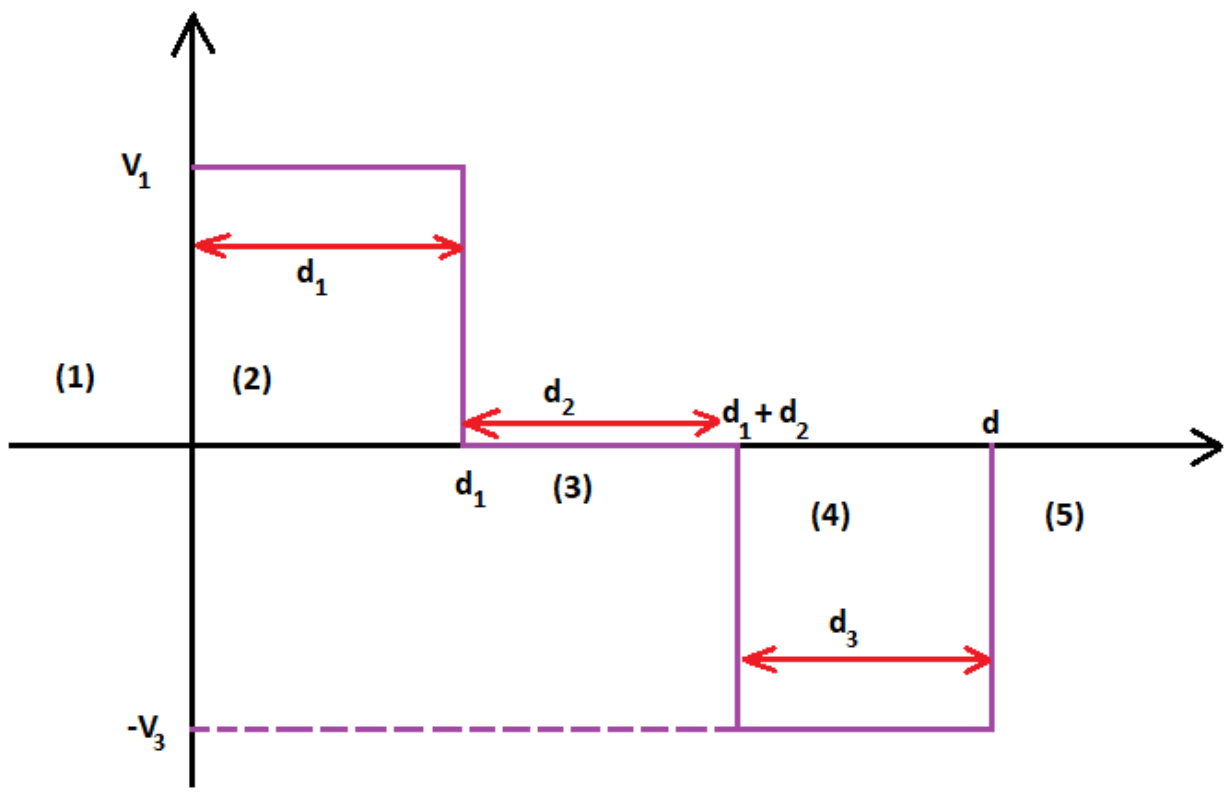}
	\caption{{(color online) Diagram illustrating the potential double-barrier  $V(x)$ that interacts with fermions in phosphorene. It has two heights $(V_1,0,V_3)$, each corresponding to a distinct region $j$ of width $d_j$, with the total width defined by $d=(d_1+d_2+d_3)$.}}\label{f01}
\end{figure}

%%%%%%%%%%%%%%%%%%%%%%%%%%%%%%%%%%%%%%%%%%%%%%%%%%%%%
%
We study Dirac fermions in phosphorene interacting with a potential double potentials (barrierm well), focusing on the analysis of the GH shifts of the transmitted beam. The system is divided into 5 regions labeled by the index $j = 1, \cdots, 5$, as shown in Fig. \text{\ref{f01}}. The incident charged particle beam with energy $E$ and incident angle $\phi$ comes from the input region $(j = 1)$, at the far left in Fig. \text{\ref{f01}}. The transmitted beam is then detected in the output region $(j = 5)$ with a lateral shift $S_t$, called the GH shift. For our purposes, we fix the double potentials $V_j(x)$ as follows
\begin{equation}
	\label{eq 1}
	V_j(x)=\left\{\begin{array}{llll}
		{V_1=V}, \qquad&  \qquad{0<x<d_1} \\
		{V_2=0},\qquad &  \qquad{d_1<x<d_1+d_2} \\
		{V_3=-V},\qquad &   \qquad{d_1+d_2 < x  < d } \\
		{0}, \qquad & \qquad\mbox{otherwise}.
	\end{array}\right.
\end{equation} 
The width of the scattering region is defined as $d = d_1 + d_2 + d_3$, {with $d_1=d_3=\frac{1-q_2}{2}d$ and $d_2=q_2d$}, see Fig. \text{\ref{f01}}.  This sets up the framework for the theoretical analysis to follow.

{The continuum approximation is well-suited to describe the physics of phosphorene, as shown by recent work using the five-hopping parameter technique \cite{ref90,ref2}, which gives remarkably accurate results within its validation limit. The monolayer Hamiltonian can be rewritten in a simpler block form using a unitary transformation. We examine the block Hamiltonian provided by \cite{ref90} 
	\begin{equation}
		H^\pm(k)=\begin{pmatrix}
			t_{AA}(k) \pm t_{AD}(k) & 	t_{AB}(k) \pm t_{AC}(k) \\
			t_{AB}^*(k) \pm t_{AC}^*(k) & t_{AA}(k) \pm t_{AD}(k) \\
		\end{pmatrix}
\end{equation}
where the second order expansion of the structural factors around the long wavelength $k = 0$ ($\Gamma$ point) allows us to write them as
\begin{align}
	&	t_{AA}(k)=\delta_{AA}+\eta_{AA} k_{y}^{2}+\gamma_{AA} k_{x}^{2}  \\
	&
	t_{AB}(k)=\delta_{AB}+\eta_{AB} k_{y}^{2}+\gamma_{AB} k_{x}^{2}+i\chi_{AB} k_x
	\\
	&
	t_{AC}(k)=\delta_{AC}+\eta_{AC} k_{y}^{2}+\gamma_{AC} k_{x}^{2}+i\chi_{AC} k_x 
	\\
	&
	t_{AD}(k)=\delta_{AD}+\eta_{AD} k_{y}^{2}+\gamma_{AD}.  
\end{align}
Now by setting the following parameters
\begin{align}
	&u_{0}=\delta_{AA}+ \delta_{AD},\qquad  \delta=\delta_{AB}+ \delta_{AC}\\
	&\eta_{x}=\eta_{AA}+ \eta_{AD}, \qquad \eta_{y}=\gamma_{AA}+ \gamma_{AD}\\ &\gamma_{x}=\eta_{AB}+\eta_{AC}, \qquad \gamma_{y}=\gamma_{AB}+\gamma_{AC}\\
	& 
	\chi=\chi_{AB}+ \chi_{AC}
\end{align}
%By developing the effective Hamiltonian around the high symmetry point $k = 0$ ($\Gamma$ point) and keeping only second order terms in $k$ and , it is shown that 
we can show that the long wavelength approximation for the Hamiltonian describing monolayer phosphorene can be expressed as \cite{ref90,ref91,ref92}}
%\begin{widetext}
	\begin{equation}\label{E2}
	H=
	\begin{pmatrix}
		u_{0}+ \eta_{x} k_{x}^{2}+\eta_{y} k_{y}^{2} & \delta + \gamma_{x} k_{x}^{2}+\gamma_{y} k_{y}^{2}+i\chi k_x \\
		\delta+ \gamma_{x} k_{x}^{2}+\gamma_{y} k_{y}^{2}-i\chi k_x & u_{0}+ \eta_{x} k_{x}^{2}+\eta_{y} k_{y}^{2}\\
	\end{pmatrix}%
\end{equation}
%\end{widetext}
with ${u}_0=-0.42$ {eV}, $\eta_{{x}}=0.58$ {eV} \AA$^2$, $\eta_{{y}}=1.01$ {eV}  \AA$^2$, $\delta=0.76$ {eV}, $\chi=5.25$ {eV} \AA, $\gamma_{{x}}=3.93$ {eV}  \AA$^2$ and $\gamma_{{y}}=3.83$ {eV} \AA.
The combination of the potential barrier \eqref{eq 1} and the effective Hamiltonian \eqref{E2} allows us to formulate the wave equation within each distinct region \(j\) using the total Hamiltonian
%We combine the scattering potential \eqref{eq 1} and the effective Hamiltonian \eqref{E2} to write our wave equation in each region $j$ using the total Hamiltonian
\begin{equation}\label{E3}
	H_{j}=H+V_{j}(x)\mathbb{I}_2
\end{equation}
%%%%%%%%%%%%%%%%%%%%%
where $\mathbb{I}_2$ is the unit matrix. 
This process facilitates a comprehensive understanding of the dynamics of the system across different regions, integrating both scattering effects and the overall system dynamics. 

Using the eigenvalue equation $~H_{j}\Psi_j = E_j \Psi_j$, which is satisfied by the spinor $\Psi_j$, one can show that the dispersion relation has the form
%\begin{widetext}
	\begin{align}\label{E4}
	E_{j}=& V_j+u_{0}+\eta_{x}k_{j}^{2}+\eta_{y}k_{y}^{2}\\
	&+s_{j}\sqrt{\left(\delta+ \gamma_{x} k_{j}^{2}+\gamma_{y} k_{y}^{2}\right)^2+\left(\chi k_{j}\right)^2}
	\nonumber
\end{align}
%\end{widetext}
with $s_{j}={\mbox{sgn}}{\left(E-V_j-u_{0}-\eta_{x}k_{x}^{2}-\eta_{y}k_{y}^{2}\right)}$. 
Given the $u_{0},\delta,k_{j}\gg k_{j}^{2}$ conditions at low momenta, we can simplify the energy spectrum by neglecting the term proportional to $k_{j}^{2}<1$, since it is much smaller than the linear terms. This simplification implies that the longitudinal wave vector $k_j$ for region $j$ has the form
%
%Using the fact that $u_{0},\delta,k_{j}\gg k_{j}^{2}$ at low momenta, the term proportional to  $ k_{j}^{2}<1$  could be neglected in the energy spectrum. The  are provided by
\begin{align}
	k_{j}=\chi^{-1}\sqrt{\left(E-V_j-u_{0}-\eta_{y}k_{y}^{2}\right)^2+\left(\delta+\gamma_{y} k_{y}^{2}\right)^2}.\label{E5}
\end{align}
Taking into account the conservation of the transverse wave vector \(k_y\), we can express the spinors of the carriers moving along the \(\pm x\) directions as \(\Psi_j(x, y) = \phi_j(x) e^{ik_y y}\). In the first region \((j = 1, x < 0)\) we use the eigenvalue equation to show that the corresponding spinor is given by a linear combination of incident and reflected waves, which can be expressed as follows:
%
%Taking into consideration the conservation of the transverse wave vector, $k_y$, we can express the spinors of the carriers moving along the $\pm x$-directions as $\Psi_j(x, y) =\phi_j(x)e^{ik_yy}$. Then, in the first region $(j=1, x<0)$ we use the eigenvalue equation to show that the corresponding spinor is given by a linear combination of incident and reflected waves  is given by
\begin{align}\label{E6}
\phi_{1}(x)=
\begin{pmatrix}
e^{i k_{1}x} & e^{-i k_{1}x}\\
z_{1}e^{i k_{1}x} & z_{1}^{-1}e^{-i k_{1}x}
\end{pmatrix}
\dbinom
		{1} 
		{r}= L_{1}[x]
	\dbinom
		{1} 
		{r}
%	\phi_{1}(x)&= \dbinom{1}{z_{1}} e^{i k_{1}x} + r
%	\dbinom
%		{1} 
%		{z_{1}^{-1}}  e^{-ik_{1}x}\\
%		&=L_{1}[x]
%	\dbinom
%		{1} 
%		{r}.
\end{align}
where $r$ is the refelction coefficient.
For regions {$j= 2, 3,$ and $4$},  the corresponding solutions are
given by 
\begin{align}\label{E7}
	\phi_{j}(x)=\begin{pmatrix}
e^{i k_{j}x} & e^{-i k_{j}x}\\
z_{1}e^{i k_{j}x} & z_{1}^{-1}e^{-i k_{j}x}
\end{pmatrix}
\dbinom
		{1} 
		{r}= L_{j}[x]
	\dbinom
		{a_j} 
		{b_j}
%	
%	
%	&=  a_j\dbinom
%		{1} 
%		{z_{j}}  e^{i k_{j}x} + b_j\dbinom
%		{1} 
%		{z_{j}^{-1}}  e^{-ik_{j}x}\\
%		&=L_{j}[x]\dbinom
%		{a_j} 
%		{b_j} 
\end{align}
with ${a_j} $ and ${b_j}$ are two constant.
For region $j=5$, the spinor can be obtained as
\begin{align}\label{E8}
	\phi_{5}(x)=  \begin{pmatrix}
1 & 0\\
z_{1} & 0
\end{pmatrix} e^{i k_{1}x}
\dbinom
		{t} 
		{0}= L_{5}[x]
	\dbinom
		{t} 
		{0}.
%	
%	t \dbinom
%		{1} 
%		{z_{1}} 
%	 e^{i k_{1}x} \\
%	 &=L_{1}[x]\dbinom
%		{t} 
%		{0} 
\end{align}
In the above equations, each complex number $z_{j}$ has the following form
\begin{align}
	z_{j}
=s_{j}e^{-i\alpha_{j}}, \quad
\alpha_{j}=\arctan\left( \frac{\chi k_{j}}{\delta+\gamma_{y} k_{y}^{2}}\right).
\end{align}
We will explore how the solutions to the energy spectrum can be used to address various questions related to transport properties. Specifically, the transmission probability and the corresponding GH shifts will be investigated.

%%%%%%%%%%%%%%%%%%%%%%%
%%%%%%%%%%%%%%%%%%%%
%%%%%%%%%%%%%%%%%%%

\section{Goss-Hänchen shifts }\label{Sec3}
%%%%%%%%%%%%%%%%%%%%%%%%%%%%%%%%%%%
Using the continuity of the eigenspinors at the interfaces \((x = x_1 = 0, x_2 = d_1, x_3 = d_1 + d_2, x = d \text{ with } d = d_1 + d_2 + d_3)\), we first calculate the reflection and transmission amplitudes, leading to the GH shifts. This procedure results in 
%%%%%%%%%%%%%%%%%%%%%
\begin{widetext}
	\begin{align}\label{E9}
	\dbinom
		{1} 
		{r} =
	{L_{1}}^{-1}[0]\ L_{2}[0]\ {L_{2}}^{-1}[d_{1}]\ L_{3}[d_{1}]\ {L_{3}}^{-1}[d_{1}+d_{2}]\ L_{4}[d_{1}+d_{2}]\ {L_{4}}^{-1}[d] \ L_{5}[d]\
	\dbinom
		{t} 
		{0}
		=\mathcal{M}
	\dbinom
		{t} 
		{0}
\end{align}
\end{widetext}
where the transfer matrix has the form
\begin{align}\label{E10}
	\mathcal{M}=\begin{pmatrix}
		\mathcal{M}_{11} &	\mathcal{M}_{12} \\
			\mathcal{M}_{21} &	\mathcal{M}_{22}\\
	\end{pmatrix}
\end{align}
and, as stated in \eqref{E9}, each of its elements can be retrieved from those of the matrix product between $L_j$ and their inverses.
%\begin{equation}\label{E12}
%	\dbinom
%		{1} 
%		{r} =\mathcal{M}
%	\dbinom
%		{t} 
%		{0}
%\end{equation}
%where $\mathcal{M}$ is given by
%\begin{align}\label{E10}
%	\mathcal{M}={L_{1}}^{-1}\ [0]\ Q \ {L_{5}}[d]=\begin{pmatrix}
%		\mathcal{M}_{11} &	\mathcal{M}_{12} \\
%			\mathcal{M}_{21} &	\mathcal{M}_{22}\\
%	\end{pmatrix}%
%\end{align}
%%%%%%%%%%%%%%%%%%%%%%%%%%%%%%%%%%%%%%%%%%%%%%%
%and the matrix $Q$ reads as
%%%%%%%%%%%%%%%%%%%%
%%%%%%%%%%%%%%%%%%%%%%
%%\begin{widetext}
%	\begin{align}\label{E11}
%		Q=L_{2}[0]\ {L_{2}}^{-1}[d_{1}]\ L_{3}[d_{1}]\ {L_{3}}^{-1}[d_{1}+d_{2}]\ L_{4}[d_{1}+d_{2}]\ {L_{4}}^{-1}[d].
%\end{align}
%\end{widetext}
%and we have defined $d=d_1+d_2+d_3$. 
%From the above results, we can rewrite  \eqref{E9} connecting the input and output in the  elegant manner
As a result, the transmission and reflection amplitudes are easily identified as
\begin{align} \label{E13}
t=\frac{1}{ \mathcal{M}_{11}}, \quad
%\label{E14}
	r=\frac{ \mathcal{M}_{21}}{\mathcal{M}_{11}}.
\end{align}
%For our task, it is convenient to 
%Because of need, let us calculate the inverse of the amplitude \(t\). Indeeed,  After a series of intricate algebraic steps, we eventually derive the  following expression 
%in terms of the physical parameters characterizing our system
%As it will be clear in the next, we introduce the inverse of $t$. Indeed, 
%after some tedious algebra, we end up with the form
%It is convenient for our task to find the inverse of $t$, and then, after some tedious algebra, we end up with the form
%
%takes the following form
%After some tedious algebras, we can show that the inverse of $t$ takes the following form
For necessity, let us calculate the inverse $t^{-1}$ of the transmission amplitude. Indeed, after a series of complicated algebraic steps, we finally derive the following expression:
\begin{equation}\label{E15}
	t^{-1}={\frac {i{{\rm e}^{ik_{{1}}d}}}{ 2 s_2 \sin^2  \alpha_{{1}}
			 \sin  \alpha_{{2}}  \sin
			\alpha_{{4}} }}\left(\zeta+\vartheta+\varkappa+\varsigma+\tau+ \beta \right) 
\end{equation}
where we have defined the quantities 
\begin{widetext}
	\begin{align}
&	\zeta=2\sin \left( d_{{1}}k_{{2}} \right) \sin \left( d_{{2}}k_{{1}}
	\right) \sin \left( d_{{3}}k_{{4}} \right)  \zeta^+\label{E16}
\\
&\label{E17}
	\zeta^+=  \sin  \alpha
	_{{1}}  \left[ is_{{1}}\cos \alpha_{{4}}  -is_{{
			2}}\cos  \alpha_{{4}}  \cos  \alpha_{{1}}  
	\cos  \alpha_{{2}}  -is_{{4}}\cos \alpha_{{1}}
	\right] -s_{{1}}s_{{2}}s_{{
			4}}\cos  \alpha_{{2}}   \left[ \cos  \alpha_{{1}}  -i\sin  \alpha_{{1}}
	\right] 
	\\
&\label{E18}
	\vartheta=	2\sin  \alpha_{{4}}  \sin \left( d_{{1}}k_{{2}}
	\right) \sin \left( d_{{2}}k_{{1}} \right) \cos \left( d_{{3}}k_{{4}}
	\right)  \left[ -2s_{{1}}\cos  \alpha_{{1}}  +s_{{2}}
 \cos^{2}  \alpha_{{1}}  \cos  
	\alpha_{{2}}  +s_{{2}}\cos  \alpha_{{2}}  
	\right]
	\\
&\label{E19}
	\varkappa=2\sin \alpha_{{1}} \sin \left( d_{{1}}k_{{2}}
	\right) \sin \left( d_{{3}}k_{{4}} \right) \cos \left( d_{{2}}k_{{1}}
	\right)  \left[ -s_{{1}}\cos  \alpha_{{4}}  +s_{{2}}\cos
	 \alpha_{{1}}  \cos \alpha_{{4}}  \cos
	 \alpha_{{2}}  -s_{{4}}\cos  \alpha_{{1}}  
	\right]
\\
&\label{E20}
	\varsigma =-2s_{{2}}\varsigma^+\sin  \alpha_{{2}}  \cos \left( d_{{1}}k_{{2}}
	\right) \sin \left( d_{{2}}k_{{1}} \right) 
\\
&\label{E21}
	\varsigma^+ = \sin \left( d_{{3}
	}k_{{4}} \right) \cos  \alpha_{{4}}   \left[ \sin^{2}
	 \alpha_{{1}} +s_{{1}}\cos  \alpha_{
		{1}}  s_{{4}} +i \cos
	 \alpha_{{1}}  \sin  \alpha_{{1}}  \cos
	\left( d_{{3}}k_{{4}} \right) \sin  \alpha_{{4}}    \right] 
\\
&\label{E22}
	\tau =-2s_{{2}}\sin  \alpha_{{1}} \cos \left( d_{{1}}k_{{2}} \right) \cos \left( d_{{2}}k_{{1}}
	\right) \sin \left( d_{{3}}k_{{4}} \right)\sin  \alpha_{{2}}
	   \left[ i\sin 
	\alpha_{{1}} \cos  \alpha_{{4}}  +s_{{1}}s_{{4}}
	\right]
\\
&\label{E23}
	\beta =-2\sin  \alpha_{{1}}  \sin  \alpha_{{4}}  
	s_{{2}}\cos \left( d_{{2}}k_{{1}} \right) \cos \left( d_{{3}}k_{{4}}
	\right)  \left[ i\sin  \alpha_{{1}}  \sin \left( d_{{1}}
	k_{{2}} \right) \cos  \alpha_{{2}}  -\cos  \alpha_{
		{1}}  \cos \left( d_{{1}}k_{{2}} \right) \sin  \alpha_{{2
	}} \right.
\end{align}
\end{widetext}

We study the GH shifts induced by the spatially modulated potential in phosphorene under certain conditions. In fact, for a given transverse wave vector $k_{y0}$ corresponding to the incident angle  $\varOmega_1(k_{y0})$, we consider the incident, reflected, and transmitted beams falling with an incident angle in the interval $[0, \pi/2]$. These beams are each represented in terms of the eigenspinors as follows
%For a given transverse wave vector $k_{y0}$ that corresponds to the incidence angle $\varOmega_1(k_{y0})$, we account for the incident, reflected, and transmitted beams that fall with an incident angle in the range $[0, \pi/2]$, in order to investigate the GH shifts induced by the spatially modulated potential in phosphorene. The incident $\Pi_i(x,y)$, reflected $\Pi_r(x,y)$, and transmitted $\Pi_r(x,y)$ beams are then given by
%\begin{widetext}
	\begin{align}\label{E24}
&	\Pi_i= \int_{-\infty}^{+\infty} g(k_y-k_{y0})e^{ik x+i k_y y}\dbinom
		{1} 
		{e^{i\varOmega(k_y)}}  \,dk_y 
\\
&\label{E25}
	\Pi_r= \int_{-\infty}^{+\infty} r(k_y)g(k_y-k_{y0})e^{-ik x+i k_y y}\dbinom
		{1} 
		{-e^{-i\varOmega(k_y)}}  \,dk_y 
\\
&\label{E26}
	\Pi_t= \int_{-\infty}^{+\infty} t(k_y)g(k_y-k_{y0})e^{ik x+i k_y y}\dbinom
		{1}
		{e^{i\varOmega(k_y)}}  \,dk_y 
\end{align}
%\end{widetext}
and $g(k_y-k_{y0})$ is a Gaussian and single peaked positive definite function around the primary wave vector $k_{y0}$. It describes the well collimated incident beam. Using boundary conditions, the reflection $r(k_y)$ and transmission $t(k_y)$ can be written in complex notation as 
\begin{align}\label{E27}
	t=\rho_t e^{i\varphi_{t}}, \quad r=\rho_r e^{i\varphi_{r}}
\end{align}
where the modulus and  phase shifts are given by 
\begin{align}\label{E28}
&	\rho_t=\sqrt{\Re^2[t]+\Im^2[t]}, \quad
 \rho_r=\sqrt{\Re^2[r]+\Im^2[r]}\\
&\label{E30}
	\varphi_t=\arctan\left({i\frac{t^*-t}{t+t^*}}\right), \quad \varphi_r=\arctan\left({i\frac{r^*-r}{r+r^*}}\right).
\end{align}
From \eqref{E28}, it is easy to derive
the following transmission $T$ and reflection $R$ probabilities 
\begin{align}\label{E29}
	T=\rho^2_t, \quad
 R=\rho^2_r.
\end{align}
Now the GH shifts can be obtained by deriving the transmission and reflection phase shifts with respect to the transverse momentum $k_y$. They are
\begin{align}
	S_t=-\frac{\partial\varphi_t}{\partial k_y}	\Big|_{k_{y0}}, \quad S_r=-\frac{\partial \varphi_r}{\partial k_y}	\Big|_{k_{y0}}.
\end{align}

%Next, we will explore the aforementioned findings to examine the numerical results for both the transmission and GH shifts. 
%The transmission GH shifts is obviously related to variations of transmission phase with respect to the incidence angle $\phi$ while the reflection GH shifts is related to variations of reflection phase with respect to the incidence angle $\phi$. Since both GH shifts have similar behavior, we have decided to consider only the transmission GH shifts in this work. Specifically, our focus will be on investigating various transmission channels and their responses under different conditions related to various system physical parameters. Additionally, we will highlight the novel aspects of our findings as compared to existing literature.

Next, we explore the above findings to examine the numerical results for both the transmission and GH shifts. The transmission GH shift is directly related to variations in the transmission phase with respect to the incident angle \(\phi\), while the reflection GH shift is related to variations in the reflection phase with respect to the incident angle \(\phi\). Since both GH shifts exhibit similar behavior, we have decided to focus only on the transmission GH shifts in this work. Specifically, our investigation will focus on different transmission channels and their responses under different conditions, influenced by different physical parameters of the system. In addition, we will highlight the novel aspects of our findings compared to the existing literature.

\section{Numerical analysis }\label{Sec4}
%%%%%%%%%%%%%%

\begin{figure}[ht]
	\centering
	\subfloat[]{
		\centering
		\includegraphics[scale=0.6]{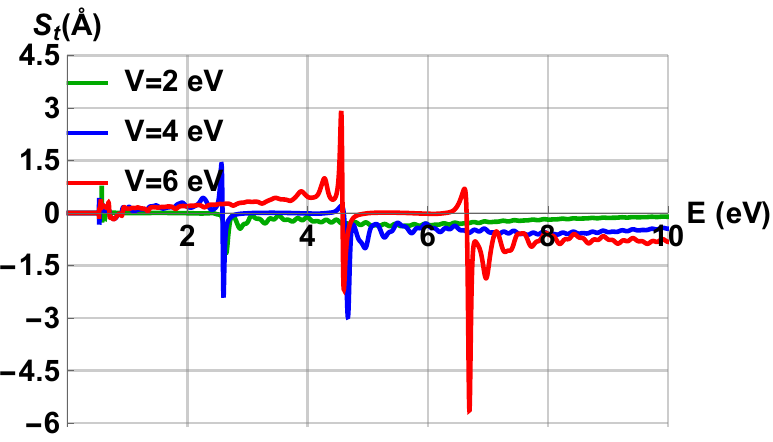}\label{f21}
	}\ \ \ \subfloat[]{
		\centering
		\includegraphics[scale=0.6]{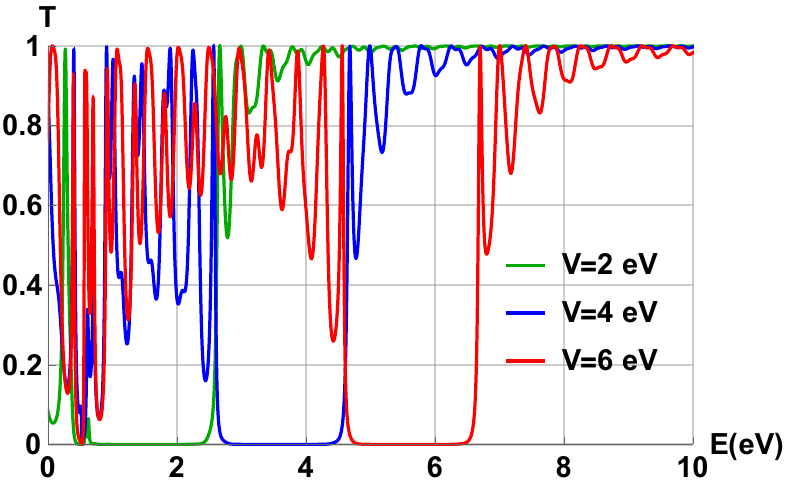}\label{f22}
	}
	\caption{The GH Shifts \textbf{\color{red}{(a)}} and transmission probability \textbf{\color{red}{(b)}} as a function of the incident energy $E $ for $d=10$  nm, $V_2=0$, $d_1=d_2=d_3=d/3$,  $\phi=4 ^{\circ}$, and three values of the  barrier heights $V_1=-V_3=V $.
		%=2 eV (green), $4$ eV  (blue),  $6$ eV (red).
	}\label{f2}
\end{figure}

{The GH shifts and the associated transmission probability are plotted against the incident energy \(E\) in Fig. \ref{f2}. During our numerical calculations, we varied the barrier height \(V\) for a number of values while keeping the incident angle \(\phi = 4^{\circ}\), the barrier width \(d = 10\) nm, and the ratio \(q_2=1/3\) fixed. Fig. \ref{f2} shows a similar pattern that is seen for all barrier height values, with a positive peak in the GH shifts consistently appearing before a negative valley. Interestingly, the energies at which the transmission becomes zero coincide exactly with the energies at which the GH shifts change sign. In particular, there is a high correlation between the transmission resonances that appear just before and just after the transmission gap region and the locations where the sign of the GH shifts change. Thus, we can deduce that the energy \(E\) that changes sign at precisely the transmission resonances located immediately before or after the zero transmission zone strongly influences the GH shifts. By tuning to the appropriate energy, this discovery provides a useful technique for achieving the desired sign of the GH shifts. Note that at this particular angle \(\phi = 4^{\circ}\), the GH shifts in phosphorene under  double potentials( barrier, well) has unique resonances.}

\begin{figure}[ht]
	\centering
	\subfloat[]{
		\centering
		\includegraphics[scale=0.6]{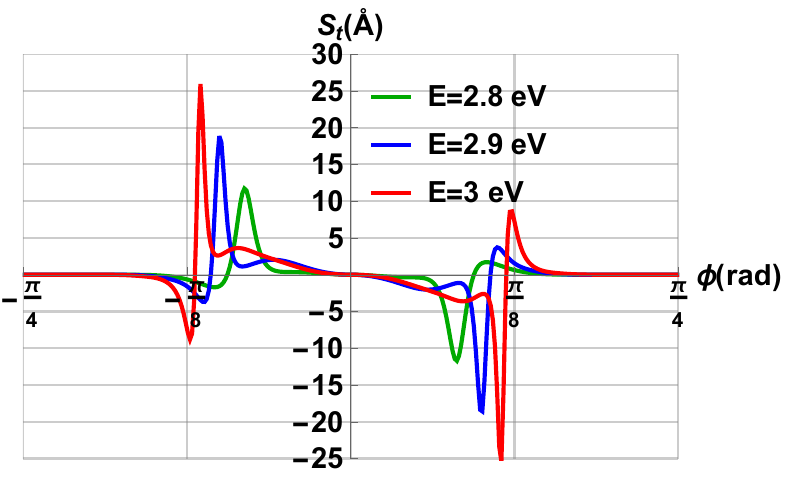}\label{f31}
	}\ \ \ \subfloat[]{
		\centering
		\includegraphics[scale=0.6]{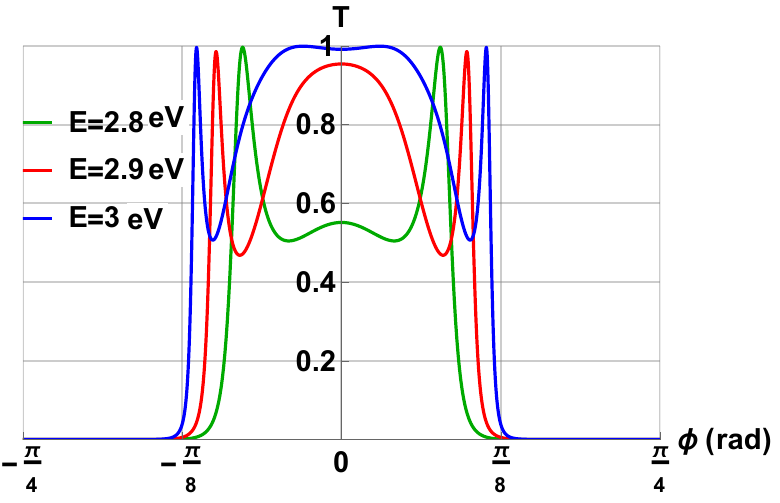}\label{f32}
	}
	\caption{The GH Shifts $S_t$ \textbf{\color{red}{(a)}}  and transmission probability \textbf{\color{red}{(b)}} as a function of the incident angle $\phi $ for $d=10$  nm, $V=2 $ eV, $d_1=d_2=d_3=d/3$,  $V_2=0$ 
		and three values of the incident  energy $E$.
	}\label{f3}\end{figure}

 Transmission and GH shifts are plotted as functions of incident angle $\phi$ in Fig. \ref{f3} for $d=10$ nm, $V=2$ eV, $q_2=1/3$, and $V_2=0$. The green, blue, and red lines correspond to incident energies $E$ = 2.8 eV, 2.9 eV, and 3 eV, respectively. Fig.~\ref{f31} shows that the GH shifts are an odd function of $\phi$, exhibiting anti-symmetric behavior around $\phi=0$. The maximum absolute values of the GH shifts increase as the incident energy $E$ is increased from 2.8 eV to 3 eV. The appearance and sign change of the GH shifts are strongly correlated with the characteristics of the transmission probability, although the GH shifts themselves are anti-symmetric in contrast to symmetric transmission. Fig.~\ref{f32} shows that the transmission probability is symmetric about normal incidence ($\phi=0$), showing behavior analogous to previous studies. However, the transmission exhibits resonances at certain incident angles, and the probability drops to zero when $\phi$ exceeds a critical value due to evanescent waves in the barrier. For incident energies $E >$ 3 eV, the maximum GH shifts are observed to decrease with increasing energy, in contrast to the trend observed for $E < 3$ eV. Contrary to graphene, the fermions in phosphorene subjected to double potentials (barrier, well) do not exhibit the Klein tunneling effect.
%%%%%%%%%%%%%%%%%%%%%%%%%%%%%%%%%%%%%%%%%

%%%%%%%%%%%%%%%%%%%%%%%%%%%%%%%%%%%%%%%%%
%%%%%%%%%%%%%%%%%%%%%%%%%%%%%%%%%%%%%%%%%

\begin{figure}[ht]
	\centering
	\subfloat[]{
		\centering
		\includegraphics[scale=0.6]{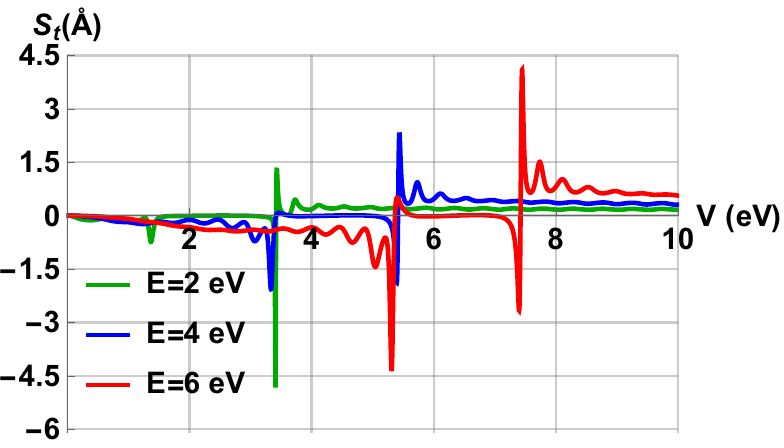}\label{f101}
	}\ \ \ \subfloat[]{
		\centering
		\includegraphics[scale=0.6]{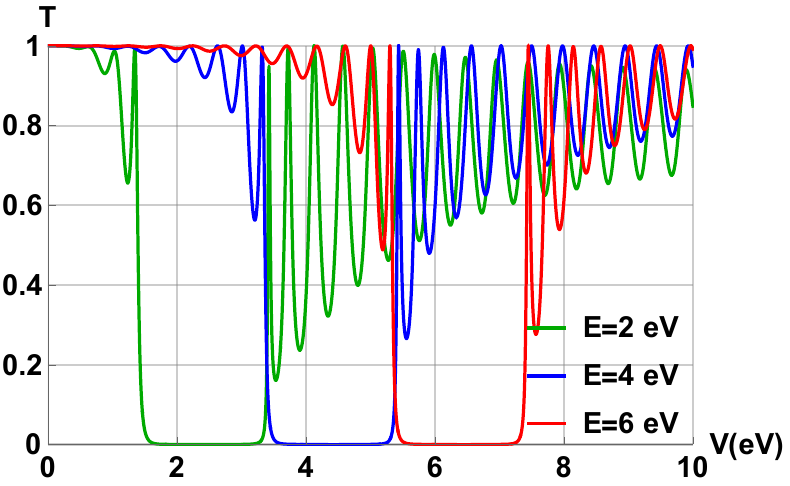}\label{f7-1}
	}
	\caption{The GH Shifts $S_t$ \textbf{\color{red}{(a)}} and transmission probability \textbf{\color{red}{(b)}} as a function of the barrier height $V$ for  $\phi= 4^{\circ}$,  $d_1=d_2=d_3=d/3$, $V_2=0 $, $d=10 $ nm, and three  values of the incident energy $E$.
	}\label{f4}
\end{figure}

In Fig. \ref{f4}, we show that the strength of the barrier height $V$ affects the GH shifts and the transmission probability. For better visualization, we have used an incident angle $\phi= 4^{\circ}$, barrier width $d=10$ nm, ratio $q_2=1/3$, and computed the transmission probabilities and the GH shifts for different values of the incident energy $E$. In contrast to what we saw in Fig. \ref{f2}, this figure shows how a negative peak precedes a positive valley in the GH shifts. This is observed for all energies. The potential values at the edges of the transmission gaps are those at which the GH shifts change sign. Thus, we can conclude that the strength of the barrier height $V$ can also be used to tune the sign of the GH shifts. This has significant implications for the potential future applications of the GH shifts in signal processing. 

%%%%%%%%%%%%%%%%%%%%%%%%%%%%%%%%%%%%%%%%%

\begin{figure}[ht]
	\centering
	\subfloat[]{
		\centering
		\includegraphics[scale=0.6]{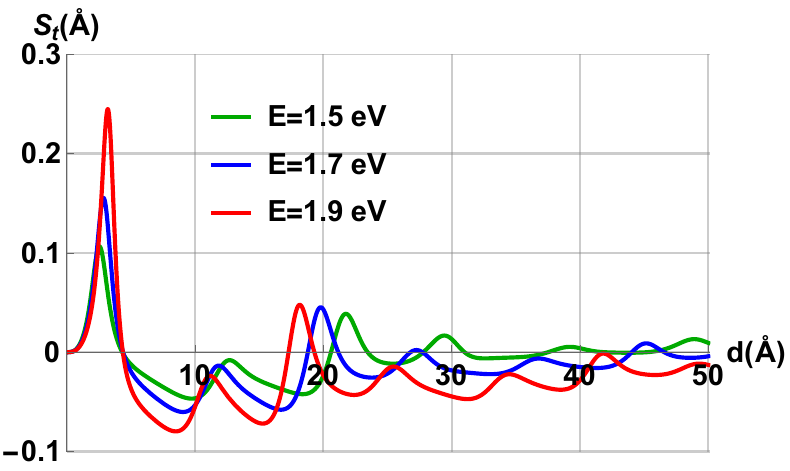}\label{f5-1}
	}
	\ \ \ \subfloat[]{
		\centering
		\includegraphics[scale=0.6]{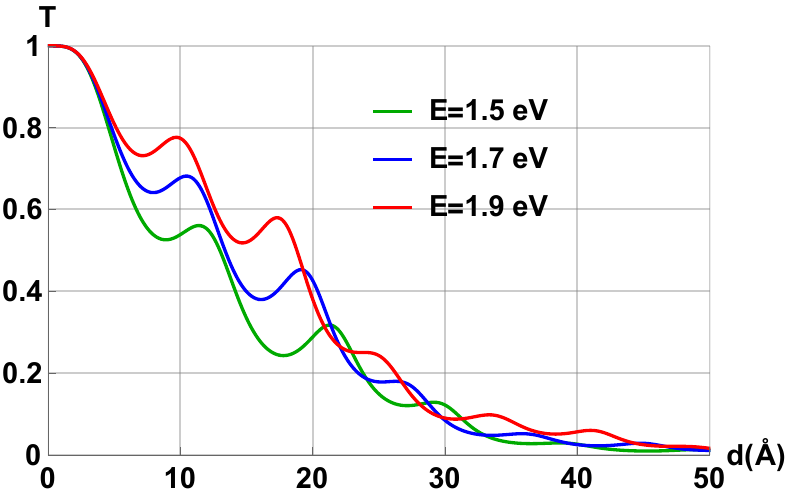}\label{f5-2}
	}
	\caption{The GH Shifts $S_t$ \textbf{\color{red}{(a)}} and transmission probability \textbf{\color{red}{(b)}} as a function of the barrier width $d$ for {$\phi=4^{\circ}$}, $d_1=d_2=d_3=d/3$,  $V_2=0 $, $V=2 $ eV, and three  values of the incident  energy $E$.}\label{f5}
\end{figure}

In Fig. \ref{f5} we show how the GH shifts and the transmission vary as a function of the full barrier width $d$ for different values of the incident energy $E$, with $\phi= 4^{\circ}$, $V=2$ eV,$V_2=0$, $q_2=1/3$. It is clear from Fig. \ref{f5-1} that the GH shifts change their sign while their magnitude increases with the incident energy $E$. We also observe that for small widths $d$ ($d<5$) the GH shifts are positive, while for ($5 < d < 15$) it becomes negative and then oscillates. Fig. \ref{f5-2} shows that the transmission decays as a function of the full barrier width $d$, which exhibits oscillatory behavior as we increase $d$, similar to what has been seen in previous work \cite{ref27}.

%%%%%%%%%%%%%%%%%%%%%%%%%%%%%%%%%%%%%%%%%
%%%%%%%%%%%%%%%%%%%%%%%%%%%%%%%%%%%%%%%%%

%%%%%%%%%%%%%%%%%%%%%%%%%%%%%%%%%%%%%%%%%

\begin{figure}[ht]
	\centering
	\subfloat[]{
		\centering
		\includegraphics[scale=0.6]{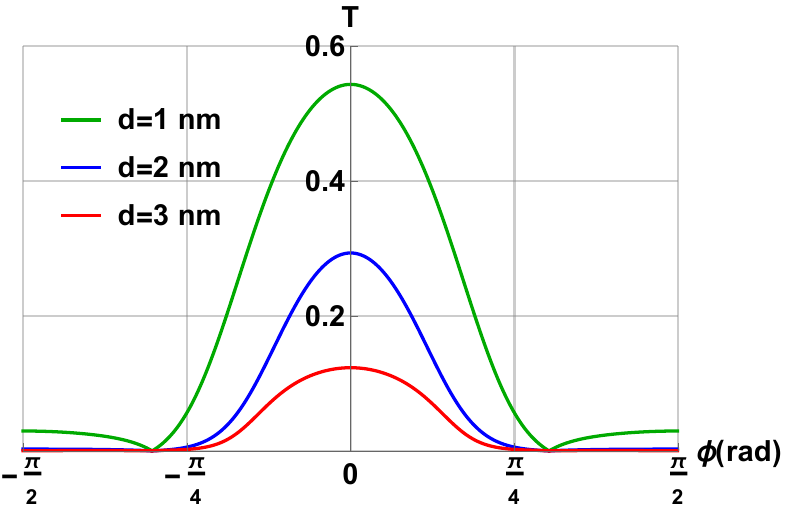}\label{f61}
	}\\
	\subfloat[]{
		\centering
		\includegraphics[scale=0.6]{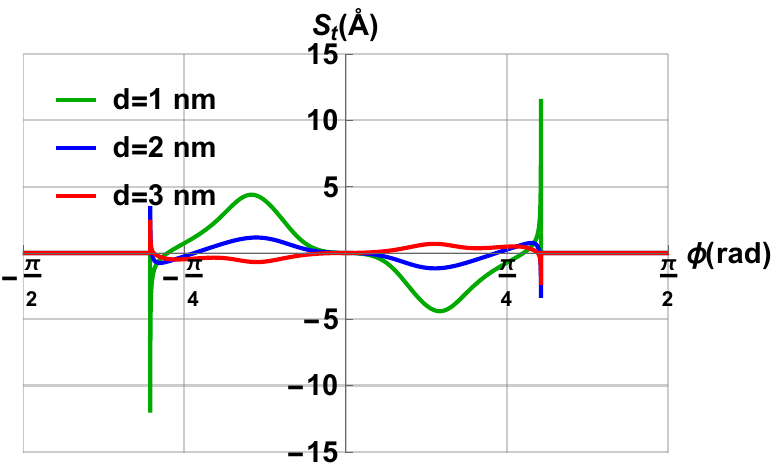}\label{f62}
	}\\
	\subfloat[]{
		\centering
		\includegraphics[scale=0.6]{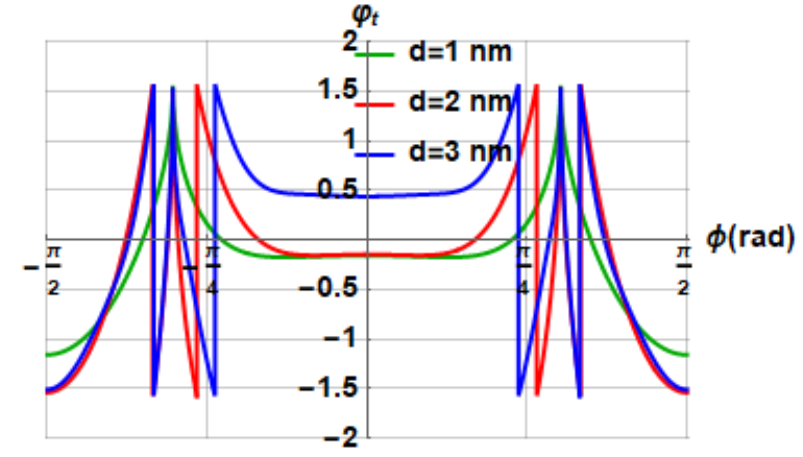}\label{f63}
	}	\caption{Transmission probability $T(\phi)$ {\color{red}{(a)}},  GH shifts $S_t(\phi)$ {\color{red}{(b)}}, and  phase shift $\varphi_t(\phi)$  {\color{red}{(c)}} as a function of  the incident angle $\phi$ for $d_1=d_2=d_3=d/3$, $V=2$ eV, $V_2=0 $, $E=1.5 $ eV, and three values of the  barrier width $d$.
		% $d=1$ (green), $d=2$ (blue), $d=3$ (red)
	}\label{f6}
\end{figure}

Fig. \ref{f6} shows the transmission, the GH shifts, and the phase shift as a function of the incident angle $\phi$, for $V=2$ eV, $V_2=0$, $q_2=1/3$, $E=1.5$ eV, and for different values of $d=1$ (green), $2$ (blue), and $3$ (red). From Fig. \ref{f61} we see that the maximum transmission occurs at normal incidence and increases with decreasing values of $d$, while it vanishes beyond some critical value of $\phi$. The transmission is symmetric with respect to $\phi=0$, as can be seen. From Fig. \ref{f62} we see that both the magnitude and the sign of the GH shifts can be changed by increasing the barrier width $d$. In Fig. \ref{f63} we see the oscillatory and symmetric behavior of the transmission phase as a function of $\phi$. This oscillatory behavior is mostly concentrated near $\phi = \pm \pi/4 $, with a broad plateau in between whose magnitude increases with the barrier width $d$.

%%%%%%%%%%%%%%%%%%%%%%%%%%%%%%%%%%%%%%%%%
%%%%%%%%%%%%%%%%%%%%%%%%%%%%%%%%%%%%%%%%%

%%%%%%%%%%%%%%%%%%%%%%%%%%%%%%%%%%%%%%%%%

%%%%%%%%%%%%%%%%%%%%%%%%%%%%%%%%%%%%%%%%%

%%%%%%%%%%%%%%%%%%%%%%%%%%%%%%%%%%%

{In summary, the study of Goose-Hänchen (GH) shifts in graphene \cite{Jellal2015,ref24,Jahani2023,JahaniPLA2023}  
	has helped to understand how Dirac fermions behave and  
	However, our work on phosphorene advances and differs from these studies in several important ways. For example, unlike graphene, phosphorene has an intrinsic direct band gap that is highly layer dependent. This is in contrast to the zero band gap of graphene, which fundamentally changes the behavior of carriers in response to external potentials. The presence of a band gap in phosphorene opens up new possibilities for controlling and tuning the GH shifts, which is not possible in gapless materials such as graphene. In addition, phosphorene is known for its anisotropic electronic properties, meaning that its behavior varies depending on the direction within the material. This anisotropy allows for more complex and nuanced control of the GH shifts compared to the isotropic nature of graphene. Our study takes advantage of this anisotropy to explore how directional dependencies affect the GH shifts, providing a level of control that is unique to phosphorene.}

%%%%%%%%%%%%%%%%%%%%%%%%%%%%%%%%%%%%%%%%%

\section{Conclusion }\label{Sec5}
%%%%%%%%%%%%%%%%%%%%%%%%%%%%%%%%%%%%%%%%%

We studied the Goos-Hänchen (GH) shifts for charged carriers scattering through  double  potential (barrierm well) in phosphorene. First, we computed the eigenvalues and eigenspinors of our Hamiltonian, which allowed us to calculate the corresponding transmission probability. Using these results, we determined the phase shifts associated with the transmission amplitude. We then computed the corresponding GH shifts and studied their behaviors with respect to various physical parameters, including the width and height of the double potentials, the incident energy, and the incident angle. {We then discussed various numerical results and highlighted their significance  of the GH shifts.}

As a result, we observed that the GH shifts are strongly dependent on the incident energy $E$. We have also pointed out that the energies at which the GH shifts reach their maximum magnitude and sign change occur at the transmission peaks just before and just after the gap region where the transmission vanishes. It is also noteworthy that the transmission resonances are highly correlated with the oscillatory behavior of the GH shifts. The transmission was found to be bilaterally symmetric with respect to the angle of incidence $\phi$, while the GH shifts where found to be an odd function of the incident angle. However, when $\phi$ exceeds a certain critical value, the transmission vanishes due to the appearance of an evanescent wave in the barrier. We also pointed out that the barrier height $V$ can also be used to control the sign of the GH shifts. 

In summary, our results confirm that the GH shifts depend strongly on the properties of the two potentials (barrier, well) and the incident angle $\phi$. Most importantly, these shifts are substantial and easily detectable, and in fact exceed those previously observed in graphene, placing phosphorene at the forefront of all future potential applications for optical sensors and switches.  Recent studies have highlighted the advantages of phosphorene in sensing applications. For instance, the high sensitivity of phosphorene to changes in its environment makes it an attractive candidate for optical sensors. Phosphorene exhibits remarkable photodetection capabilities, which can be attributed to its tunable bandgap and strong light-matter interaction, enhancing its performance in light sensing applications \cite{Cho2017}.

Moreover, note that  graphene has been extensively studied for its sensing properties \cite{JahaniMSE2023}. However, it often suffers from limitations such as low sensitivity to certain analytes then puts forward the difficulties encountered in its  application  in sensors \cite{Han2020}.
 In contrast, phosphorene's unique electronic structure allows for enhanced interactions with a wider range of molecules, making it more effective in specific sensing scenarios. As evidenced by recent work in the field, phosphorene not only shows potential for ultra-sensitive detection but also offers advantages in terms of operational flexibility and integration into existing technologies \cite{Liu2022}.

{Finally, it should be noted that GH shifts in phosphorene can be determined experimentally using a variety of electron-based \cite{Kittel2004} and optical \cite{Born1999} techniques, such as electron beam and optical interferometry. These techniques make it possible to see both positive and negative shifts, which provides insight into how the material behaves in different situations and supports the theoretical predictions we developed for our investigation.}

%%%%%%%%%%%%%%%%%%%%%%%%%%%%%%%%%%%%%%%%%
\section*{Author Contributions}

All authors contributed equally to this work. All authors have read and
approved the published version of the manuscript.

\section*{Declaration of competing interest}

The authors declare that they have no known competing financial interests or
personal relationships that might appear to influence the work presented in
this paper.

\section*{Data Availability Statement}
This manuscript has no
associated data or the data will not be deposited. [Authors’
comment: The data that support the findings of this study
are available on request from the corresponding author].
%%%%%%%%%%%%%%%%%%%%%%%%%%%%%%%%%%%%%%%%%

\end{document}